# NUMERICAL ANALYSIS OF A BEAM-ENERGY-SPREAD MONITOR USING MULTI-STRIPLINE ELECTRODES

T. Suwada[*], High Energy Accelerator Research Organization (KEK)
1-1 Oho, Tsukuba, Ibaraki 305-0801, Japan

*Abstract*

A beam-energy-spread monitor is under development in order to control and stabilize the energy spread of high-current single-bunch electron beams for producing a sufficient number of positrons. The author has proposed a new monitor using multi-stripline electrodes in order to reinforce this purpose. This report describes the basic design of the monitor based on a numerical analysis. The analysis result shows that the resolution of an energy-spread measurement is expected to be less than 0.3% for nominal operation conditions.

## 1 INTRODUCTION

The KEK B-Factory (KEKB) project[1] is progressing in order to test CP violation in the decay of B mesons. KEKB is an asymmetric electron-positron collider comprising 3.5-GeV positron and 8-GeV electron rings. The KEKB injector linac[2] was upgraded in order to inject single-bunch positron and electron beams directly into the KEKB rings. The beam charges are designed to be 0.64 nC/bunch and 1.3 nC/bunch with a maximum repetition rate of 50 Hz for the positron and electron beams, respectively. High-current primary electron beams (~10 nC/bunch) are required in order to generate a sufficient number of positrons. Since the KEKB is a factory machine, a well-controlled operation of the injector linac is required for minimizing the tuning time and maximizing stable operation. Stable control of the beam positions and energies at several sectors throughout the beam-position and energy feedback systems[3] are essential in daily operation; however, the energy spread of the primary electron beams is often enlarged due to a long-term phase drift of high-power and booster klystrons. Thus, beam diagnostic and monitoring tools are required to cure the beam energy spread; furthermore those are also expected to control the longitudinal wakefields of the high-current primary electron beams pulse-by-pulse, especially at the 180-degree arc of the injector linac. A beam-energy-spread monitor (BESM) with multi-stripline electrodes is one of the very useful monitoring tools for satisfying such requirements. The monitor has been newly designed based on a numerical analysis involving a multipole analysis of the electromagnetic field generated by charged beams. The BESM detects the spread of beam sizes at large dispersion sections by detecting any variation of the electromagnetic field distribution induced in the monitor with the multi-stripline electrodes; thus, the following method can be applied to not only measuring the energy spread, but also to measuring the transverse spatial structure of the beam.

## 2 MULTIPOLE ANALYSIS

The electromagnetic field generated by relativistic charged beams inside a conducting duct is almost boosted in the transverse direction to the beam axis due to Lorentz contraction. This phenomenon shows that if the wall loss of the image charges is negligibly small, the electromagnetic coupling of the inner surface of the duct to the beams can be well treated as a two-dimensional electrostatic potential problem. Thus, any derivation of the image charges induced on the duct is simply attributed to the electrostatic potential problem on the transverse plane. For a conducting round duct, the image charges induced by a line charge can be solved as a boundary problem in which the electrostatic potential is equal on the duct[4]. The formula for the image charge density ($j$) is given by

$$j(r,\phi,R,\theta) = \frac{I(r,\phi)}{2\pi R} \frac{R^2 - r^2}{R^2 + r^2 - 2rR\cos(\theta-\phi)}, \quad (1)$$

where $(r,\phi)$ and $(R,\theta)$ are the polar coordinates of the line charge and the pickup point on the duct, respectively; $R$ is the duct radius; and $I$ is the line charge (see fig.1). The formula can also be represented by expanding in power series of $r/R$,

$$j(r,\phi,R,\theta) = \frac{I(r,\phi)}{2\pi R}\left[1 + 2\sum_{k=1}^{\infty}\left(\frac{r}{R}\right)^k \cos k(\theta-\phi)\right]. \quad (2)$$

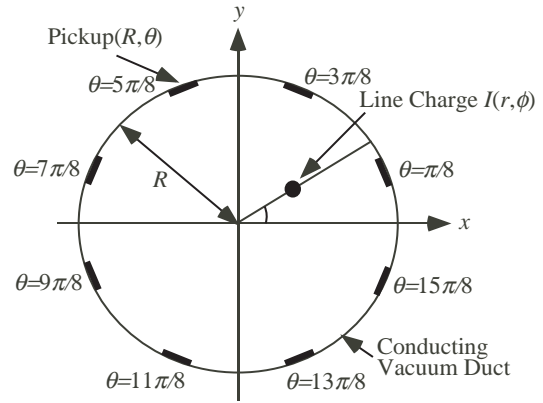

Figure 1: Polar coordinates of a line charge and eight pickups defined in a conducting duct.

---
[*] E-mail address: tsuyoshi.suwada@kek.jp

If the transverse distribution of a traveling beam is according to a Gaussian function inside the duct, the total image charge ($J$) is formulated by integrating the image charge density with a weight of the Gaussian distribution inside the duct area,

$$J(R,\theta) = \frac{I_b}{2\pi R} \iint j(r,\phi,R,\theta) \times \exp\left[\frac{-(x-x_0)^2}{2\sigma_x^2}\right]\exp\left[\frac{-(y-y_0)^2}{2\sigma_y^2}\right]dxdy, \quad (3)$$

where $I_b$ is the beam charge, $\sigma_x$ and $\sigma_y$ are the horizontal and vertical root mean square (rms) half widths of the beam, respectively, and $(x_0,y_0)$ is the charge center of the gravity of the beam. Assuming that the widths of the charge distribution are sufficiently small compared to the duct radius, $\sigma_x, \sigma_y \ll R$, integration is easily performed by extending the integration area to $x,y \to \pm\infty$, as follows:

$$J(R,\theta) \approx \frac{I_b}{2\pi R}\left\{1 + 2\left[\frac{x_0}{R}\cos\theta + \frac{y_0}{R}\sin\theta\right] + 2\left[\left(\frac{\sigma_x^2-\sigma_y^2}{R^2}+\frac{x_0^2-y_0^2}{R^2}\right)\cos 2\theta + 2\frac{x_0 y_0}{R^2}\sin 2\theta\right]\right\}, (4)$$

where the higher orders are neglected; the first to third expanded terms correspond to the monopole, dipole, quadrupole moments, respectively. The beam sizes are related to the quadrupole moment at the least order. Thus, a beam-size measurement can be performed to detect the quadrupole moment ($J_{quad}$). The formula normalized to the total image charges is given by

$$J_{quad} \equiv \int_0^{2\pi} J(R,\theta)\cos 2\theta d\theta \Big/ \int_0^{2\pi} J(R,\theta)d\theta, \quad (5)$$

$$= \left(\frac{\sigma_x^2-\sigma_y^2}{R^2}+\frac{x_0^2-y_0^2}{R^2}\right). \quad (6)$$

The quadrupole moment is dependent upon both of the beam positions and the sizes from the above formulas; however, the position dependence can be corrected by using the dipole moments. It is also approximately given by using the $n$-pickup amplitudes ($V_i$ [$i=1$-$n$]),

$$J_{quad} \approx \sum_{k=1}^n V_k \cos 2\theta_k \Big/ \sum_{k=1}^n V_k. \quad (7)$$

Normalization by summing the $n$-pickup amplitudes needs to cancel out the beam charge variation due to the beam jitter and the sensitivity dependence of the pickups due to the angular width of the electrodes. It is noticed that the absolute beam sizes can not be independently obtained, because the square difference ($\sigma_x^2 - \sigma_y^2$) of the beam sizes is only related to the quadrupole moment. This is because the equipotential lines are invariant under the condition $\sigma_x^2 - \sigma_y^2$=const if the beam positions do not change.

## 3 NUMERICAL METHOD

A numerical analysis was carried out based on a charge-simulation method[5] in order to calculate the voltages induced on the electrodes dependent on the beam position and sizes. Here, only a brief overview of this method is given (see ref.6 in detail). The method is based on the boundary element method for analyzing a two-dimensional electrostatic potential problem. In this method some boundary elements and imaginary charges are introduced in order to analyze an electrostatic-field system (see fig.2). All of the conductor surfaces in the system are divided into many domains, which are called "boundary elements"; imaginary charges are also arranged near to the boundary elements in a one-to-one manner. The electrostatic potential of each conductor can be calculated so as to satisfy the boundary conditions of the system, that is, so as that the calculated equipotential surfaces correspond to the conductor surfaces by using the linear superposition of the electrostatic field contributed by all of the real (beam) and imaginary charges. The goodness of the numerical analysis was investigated by analyzing the convergence of the electrostatic potentials calculated for each electrode in various segmentation numbers, while the transverse charge distribution obeyed a Gaussian function with a total horizontal (vertical) width of $6\sigma_x(\sigma_y)$, which was also segmented by using finite number of line charges with infinite length. The parameter $\varepsilon/\delta$ was tuned so as to produce good symmetrical and constant electrostatic potentials on the electrode surfaces. The calculation convergence was deduced to be better than 1%. The obtained parameters are summarized in Table 1.

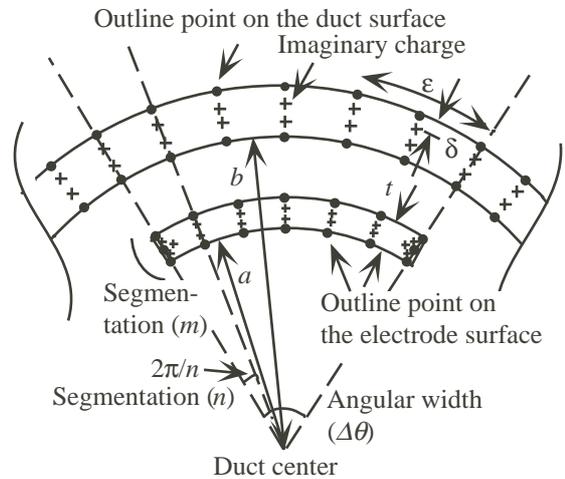

Figure 2: Segmentation of the beam duct and electrode surface based on the charge-simulation method.

Table 1: Parameters used in the numerical method

| Radial segmentation $m$ | 4 |
|---|---|
| Azimuthal segmentation $n$ | 240 |
| $\delta/\varepsilon$ | 0.6~0.7 |

## 4 APPLICATION TO THE BESM

The number of the electrodes was decided by analyzing the convergence of $\sigma_x^2 - \sigma_y^2$ using eq. (7), while the total angular width $(2/3 \times 2\pi)$ of the electrodes was constant. Figure 3 shows that good convergence was obtained at eight electrodes. The angular width should be carefully designed from a point of view in terms of the signal-to-noise ratio and the coupling strength between the electrodes[7] under the condition that the outer radius ($b$) and the electrode length ($L$) are fixed by replacing of the present beam-position monitor installed in the arc. An electrode with an angular width of 15 degree satisfies the above requirement and, furthermore, it is almost free from a direct strike on the electrode surface of which synchrotron radiation is generated at the entrance of the arc. The total angular width is $1/3 \times 2\pi$, viewed from the beam. For this angular width, the minimum detectable beam charge is estimated to be ~0.6 nC/bunch, analyzed based on the data of the present beam-position detection system[8]. The inner radius ($a$) should be determined so as to comprise a 50-$\Omega$-transmission line. Table 2 shows the design parameters of the BESM. Based on the design parameters, Fig. 4 shows that using the design parameters the variation of the quadrupole moment to the energy spread ($\sigma_E/E$) expected at the 180-degree arc, where the horizontal dispersion is $\eta_x$=7.2 mm/% and the beam energy is 1.7 GeV, with various vertical beam sizes. The beam-position dependence was neglected in this analysis for the sake of simplicity. Assuming that the noise level of the pickup amplitudes is expected to be less than 0.5%, which determines the minimum detectable charge, the detection error of the energy spread is expected to be about 0.3%.

Table 2: Design parameters of the BESM

| Inner radius $a$ [mm] | 20.6 |
|---|---|
| Outer radius $b$ [mm] | 23.4 |
| Angular width $\Delta\theta$ [deg.] | 15 |
| Electrode thickness $t$ [mm] | 1.5 |
| Stripline length $L$ [mm] | 132.5 |
| Number of electrodes | 8 |

## 5 CONCLUSIONS

A new beam-energy-spread monitor with eight-stripline electrodes was designed on the basis of the multipole analysis for charged beams. Eight electrodes with each angular width of 15 degree are required to detect the precise quadrupole moment for a beam charge greater than 0.6 nC/bunch. The rms resolution of the energy-spread measurement is expected to be 0.3% for nominal operation conditions. This is small enough to stably transport high-current primary electron beams at the 180-degree arc in the injector linac.

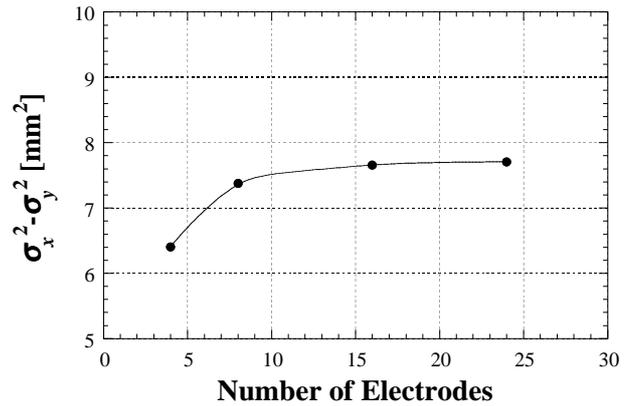

Figure 3: Variation of the square difference of the beam sizes depending upon the number of electrodes. The horizontal and vertical beam sizes are assumed to be $\sigma_y$=3 and $\sigma_y$=1 mm, respectively.

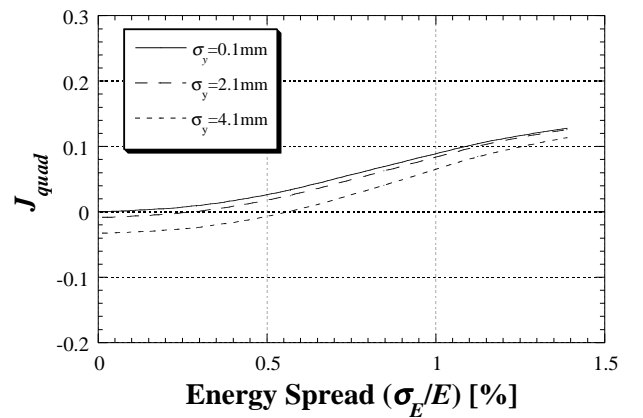

Figure 4: Variation of the quadrupole moment to the energy spread expected at the 180-degree arc.